# Observation of gate-tunable coherent perfect absorption of terahertz radiation in graphene

Nurbek Kakenov[1], Osman Balci[1], Taylan Takan[2], Vedat Ali Ozkan[2],

Hakan Altan[2], Coskun Kocabas[1]

Department of Physics, Bilkent University, 06800 Ankara, Turkey

Department of Physics, Middle East Technical University, 06800 Ankara, Turkey

**Abstract:** We report experimental observation of electrically-tunable coherent perfect absorption (CPA) of terahertz (THz) radiation in graphene. We develop a reflection-type tunable THz cavity formed by a large-area graphene layer, a metallic reflective electrode and an electrolytic medium in between. Ionic gating in the THz cavity allows us to tune the Fermi energy of graphene up to 1eV and to achieve critical coupling condition at 2.8 THz with absorption of 99%. With the enhanced THz absorption, we were able to measure the Fermi energy dependence of the transport scattering time of highly doped graphene. Furthermore, we demonstrate flexible active THz surfaces that yield large modulation in the THz reflectivity with low insertion losses. We anticipate that the gate-tunable CPA will lead efficient active THz optoelectronics applications.

The phenomena of coherent perfect absorption (CPA) is the time-reversed analog of stimulated emission [1-3]. The optical absorption of a conducting thin film, which is limited to maximum of 50% in free standing form, can be enhanced under illumination of two coherent light beams when they are in-phase on the film. The concept of CPA has been implemented to various materials systems such as, metamaterials[4], two-level atomic systems[5], phase change materials[6], plasmonic systems[7] and radar absorbing surfaces[8]. Very recently the enhancement of optical absorption in 2-dimensional conductors has attracted great attention for realization of gate-tunable optoelectronic devices. Enhancement of optical absorption in graphene, in particular, plays an important role for broadband tunable optoelectronic devices. The ability to control rates of interband[9-11] and intraband[12] electronic transitions via electrostatic gating, enables novel active optoelectronic devices. At optical wavelengths, the optical absorption in graphene is limited to 2.3%[10, 11, 13], however for longer wavelengths (THz [12, 14, 15] and microwave[8]) absorption can be increased up to 50% when the surface impedance of graphene ($Z_G$) matches the half of the free space impedance[16], $Z_G = 1/\sigma(\omega) = Z_0/2$ where $Z_0$ is the free space impedance and $\sigma(\omega)$ is the optical conductivity (See the small signal model given in Supporting Materials Fig.S1). To enhance the optical absorption further, various device structures have been explored. Pattering graphene into ribbons leads to enhanced absorption due to the localized plasmon oscillations. Fang *et al* demonstrated absorption of 20% in far-IR frequencies[17]. Placing graphene on a photonic crystal cavity[18] or inside a microcavity[19, 20] enhances the absorption due to multiple passes. Very recently, Thareja *et al* placed graphene at a quarter-wave-distance from a metallic surface and showed enhancement up to 5.5% in IR wavelengths[21-23]. With the help of local plasma frequency, complete optical absorption at IR frequencies has been proposed using periodically patterned doped graphene[24, 25].

Gate-tunable coherent absorption in graphene at terahertz frequencies has more technological importance because of being a low cost alternative material for active THz devices. The recent theoretical studies show that gating graphene near a reflective surface, would yield gate-tunable CPA for terahertz radiation [26, 27]. They predicted that, under coherent illumination, 100% of THz radiation can be absorbed by a highly doped monolayer graphene, when the Fermi energy is close to 1eV. Varying the doping level, THz absorption can be controlled efficiently by electrical means. This is a challenging requirement. Although, the static CPA in graphene for microwave[28] and visible[29, 30] spectra has been reported, due to the limitation of conventional gating schemes, the gate-tunable CPA of THz radiation in graphene has not been observed yet[12, 31]. In our previous works, we used ionic gating to control optical properties of graphene in a very broad spectrum extending from visible to microwave wavelengths[8, 13, 14, 32]. In this Letter, we demonstrate a new type of tunable THz cavity which enables us to observe gate-tunable CPA. Figure 1(a) shows the schematic drawing of our device structure. The large-area monolayer graphene is synthesized by chemical vapor deposition on copper foils and then transferred on 20-µm-thick porous polyethylene membrane (42% porosity) which is placed on a reflective gold electrode. The thickness of the membrane defines the cavity length and the resonance wavelength. The gold electrode operates both as the back reflecting mirror and the gate electrode. We soaked the membrane with room temperature ionic liquid ((Diethylmethyl(2-methoxyethyl) ammonium bis(trifluoromethylsulfonyl)imide, [deme][Tf2N])) which has large electrochemical window that yields tunable Fermi energy on graphene up to 1 eV. Both electrolyte and PE membrane are transparent between 0.1-15 THz (Supporting Materials Fig.S2). Figure 1(b) shows a schematic cross-sectional view of the device under a bias voltage that polarizes the ionic liquid in the membrane and forms electrical double layers (EDL) near the graphene and gold interface. The EDL electrostatically dopes the graphene layer and alters its conductivity. Since the thickness of EDL is very thin for ionic liquids, this configuration yields very large electric field and induced charges on the surface. The advantage of this device is that it provides a very efficient gating scheme with charge density up to $10^{14}$ cm$^{-2}$ and Fermi energy of 1 eV of open graphene surface. These doping levels are enough to satisfy the CPA condition at THz frequencies. Our device yields a single channel CPA, when the incident and reflected THz beams are in phase at the graphene interface. For our device structure, the resonance condition can be written as $t cos(\theta) = (2m + 1)\lambda/4n$ where $\theta$ is the incidence angle, $t$ is the thickness of the membrane, $m$ is an integer and $n$ is the index of refraction of the cavity. Spectroscopic measurements provide the resonances and anti-resonances which yield perfect and no absorption conditions, respectively.

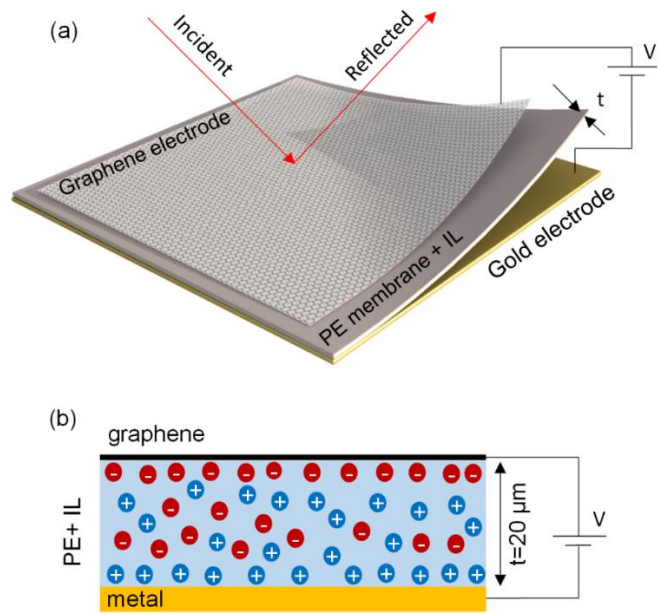

**Figure 1. Active THz surfaces**: (a) Schematic representation of electrically tunable THz cavity used for the coherent perfect absorption in graphene. The THz cavity is formed by a porous membrane sandwiched between graphene and gold electrodes. The thickness of the membrane is 20 μm. The ionic liquid electrolyte is soaked into the membrane. (b) Cross-sectional view of the cavity showing the formation of electrical double layers on the graphene and gold electrodes.

Figure 2a shows the fabricated device. We measured THz reflection from the biased device using a Fourier transform infrared spectrometer (FTIR) equipped with Far-IR detector and a Far-IR source (Figure 2b). Since ionic liquids have very low vapor pressure, we recorded the reflection spectrum under the vacuum (10 mTorr) to remove the absorption of water. Figure 2c shows the measured reflectivity spectrum from the device under different bias voltages. For the membrane thickness of 20 μm and incidence angle of 30°, we observed multiple resonance absorptions at 2.83, 8.24, and 13.23 THz frequencies. For the first resonance, we obtained absorption of 99 % at 2.0 V bias voltage. Unlike the condition of the freestanding film, CPA occurs when the real part of the optical conductivity of doped graphene reaches the values $\sigma(\omega) = 1/Z_0$ where $Z_0$ is the free space impedance. The optical conductivity of graphene in THz frequencies can be described with Drude response as

$$\sigma(\omega) = \frac{e^2}{\pi\hbar} \frac{iE_F}{\omega + i\tau^{-1}}$$

where, $E_F$ is the Fermi energy, $\tau$ is the transport scattering time. For high doping levels, $\tau$ varies with the Fermi energy. We observed the perfect absorption at low THz frequencies (< 5 THz). For higher frequencies however, the required doping levels exceed the accessible levels with the present device. The variation of the resonance reflectivity of the first three resonances are plotted in Figure 2d against the bias voltage. The reflectivity is normalized by the reflection at the charge neutrality point (CNP, around -1 V). The large shift in the CNP is associated with the work function difference between the graphene and gold electrodes. We obtained 99, 76, and 42% absorption for 2.83, 8.24, and 13.23 THz frequencies respectively. To observe perfect absorption for higher order modes, we need larger voltages which exceed the electrochemical window of the electrolyte and introduce irreversible damage on graphene electrode.

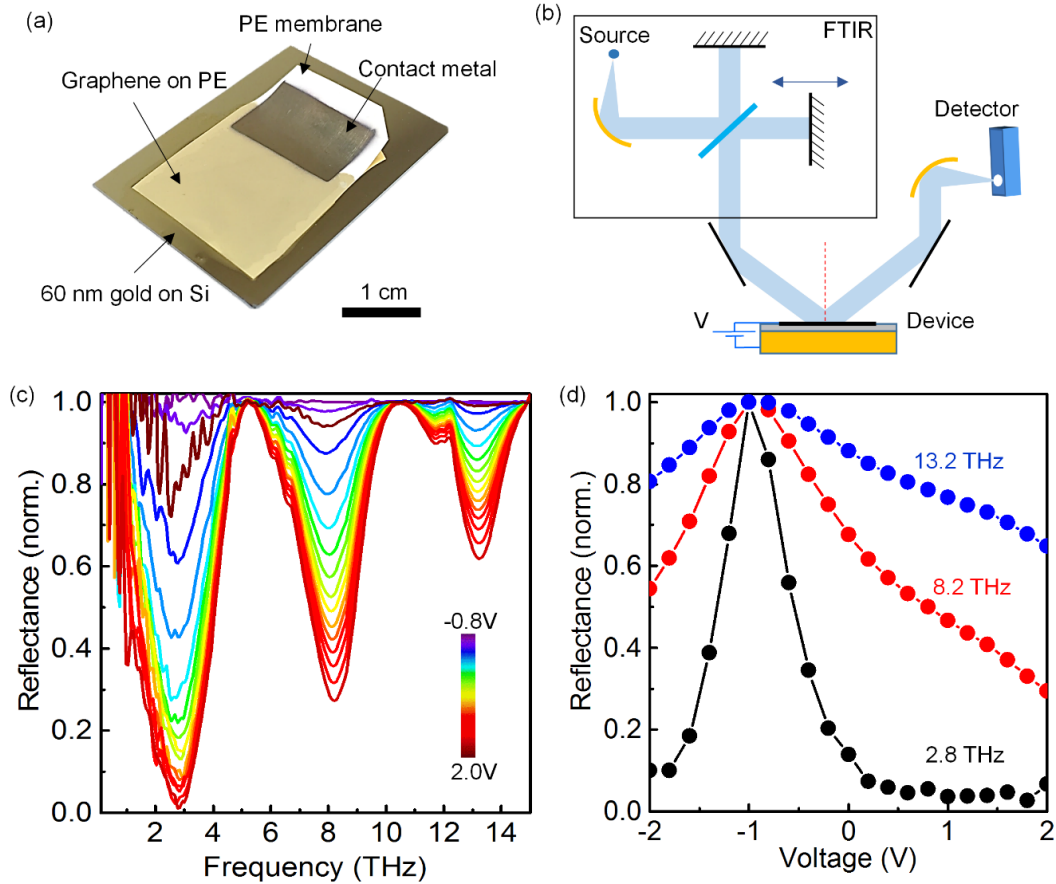

**Figure 2 Coherent perfect absorption of THz radiation**: (a) Photograph of the fabricated THz cavity. The monolayer graphene is transferred on PE membrane and placed on a gold coated substrate. The 20-μm-thick membrane defines the cavity length, holds the electrolyte and forms the mechanical support graphene. (b) Experimental setup used for the THz measurements. (c) The Reflectivity spectrum from the device at different bias voltages. (d) The variation of the resonance reflectance with gate voltage. The charge neutrality point is at -1 V.

The Fermi energy provides wealth of information about the electrical and optical properties 0f the device. Liu et al, predicted that, to achieve CPA in THz, the Fermi energy of graphene should be close to 1eV that yields the required optical conductance for the critical coupling. Near-IR and IR (see Supporting Materials Figure S4) reflection spectra from the device provide direct measurement of the Fermi energy of the doped graphene. Figure 3(a) shows the electronic band structure of doped graphene. Due to the Pauli blocking, doped graphene has a gap in the optical absorption for photon energies $E < 2E_F$. Gating graphene results an increase in the absorption gap and a step-like change in the reflectivity spectrum. Figure 3b shows the measured reflectivity spectra which show a step-like change in the reflectivity with a cutoff wavelength at $2E_F$. Although, a monolayer graphene absorbs around 1.8% on a dielectric substrate, in our cavity structure, the reflectivity shows about 3% modulation due to multiple passes. Figure 3c shows the extracted Fermi energy as a function of bias voltage. At charge neutrality point ($V_{CNP} = -1$ V) the unintentional doping level is 0.2 eV and increases linearly with the gate voltage up to 1 eV. At $V_G=0$ V, graphene is significantly doped with Fermi energy of 0.55 eV due to work function difference between the gold and graphene electrodes. To get more insight, we performed electrical characterization of the device using an LRC meter. Figure 3d shows the variation of the resistance and capacitance of the devices with the bias voltage. At charge neutrality point, the sheet resistance reaches up to 4.5 kΩ and decreases down to 0.8 kΩ which also includes the contact resistance of the electrodes. The capacitance of the device shows a minima (0.8 µF/cm$^2$) at the charge neutrality point due to the minimum quantum capacitance of the graphene layer. The electrical characterization shows a good agreement with the spectroscopic measurements. Our results suggest that the critical coupling condition is achieved when the Fermi energy is around 1eV.

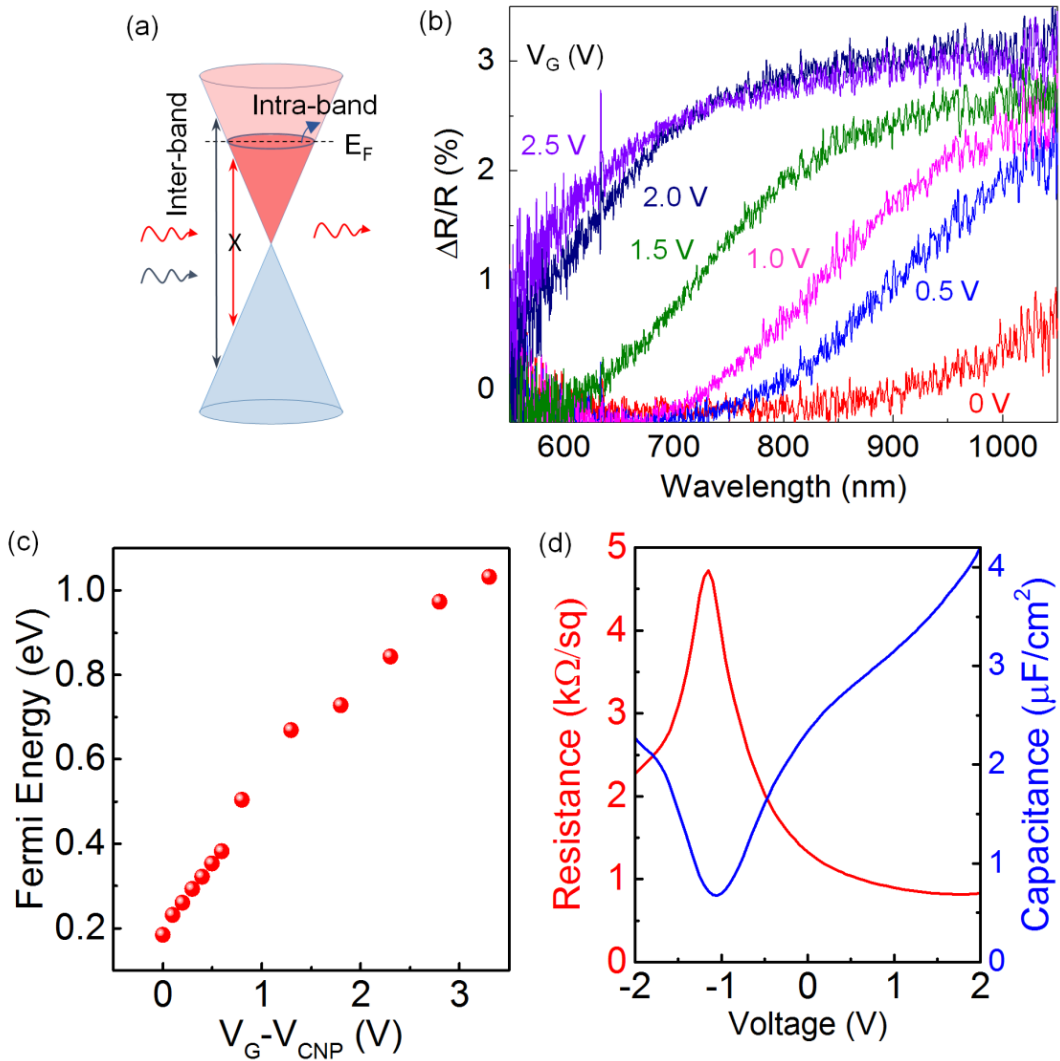

**Figure 3. Electrical and optical characterization of the device:** (a), Schematic representation of the band structure of graphene and possible electronic transitions. (b) Gate-tunable near-IR optical reflection from the graphene surface at different bias voltages. The number on the curves shows the bias voltage. (c) Fermi energy extracted from the reflection spectrum. (d) Variation of the resistance and capacitance of the devices with the bias voltage. At charge neutrality point, resistance reaches a maxima of 4.5 kΩ and the capacitance goes to a minima of 0.8 µF/cm$^2$.

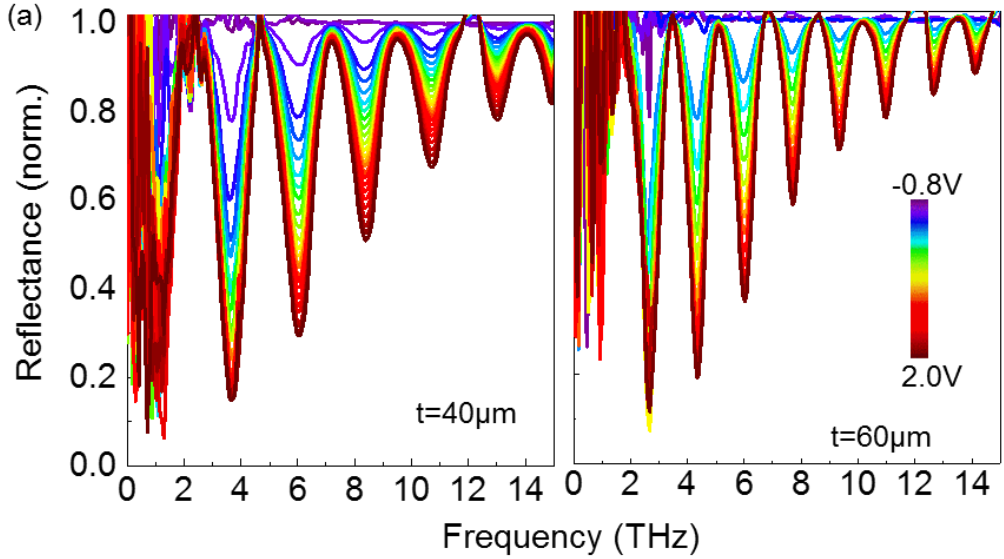

**Figure 4. Tunable reflectivity from various THz cavities**: Reflectance spectrum of THz cavities with 40 and 60 μm membrane thickness.

The thickness of the porous substrates and the incidence angle define the frequency of the resonance absorption. We repeat our measurements with different membrane thicknesses. Figure 4 shows the gate tunable reflectivity spectrum from two different devices with 40 and 60 μm cavity length. The observed resonance wavelengths satisfy the critical coupling condition as $\lambda_m = 4nt\cos(\theta)/(2m + 1)$. We don't observe a significant change in the frequency however, the width of the resonance varies slightly with the bias voltage (see Supporting Materials Fig. S4). The fundamental resonances of the large cavities are buried under the noise level, due to the sensitivity of the FTIR system at low frequencies (< 2 THz). We performed additional experiments using continuous wave tunable frequency THz sources (see Supporting Materials Fig.S5). Similarly, we obtained 98% modulation at 0.368 THz.

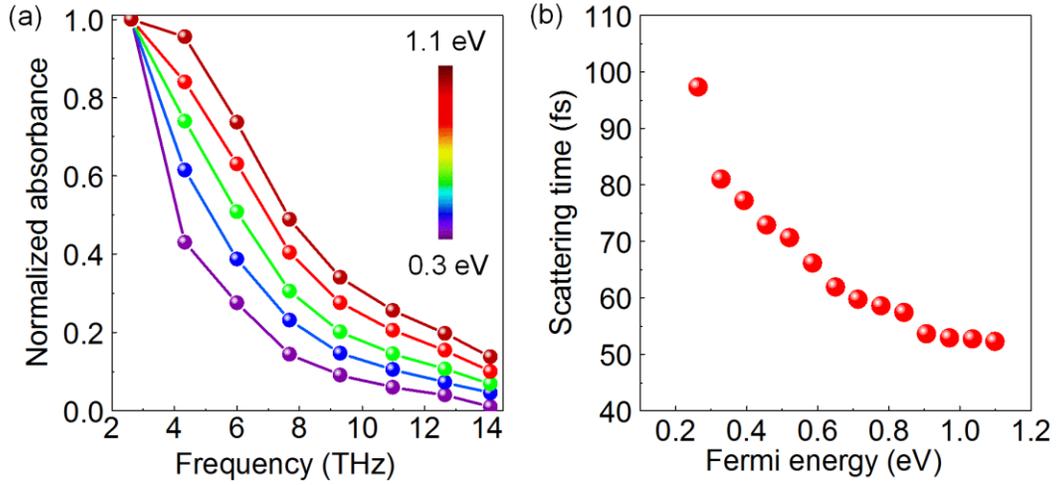

**Figure 5. Drude response of graphene at high doping level**: (a) Frequency dependence of the resonance absorbance at different Fermi energies. The absorbance is normalized by maximum absorbance at 2.8 THz. (b) The variation of the transport scattering time with the Fermi energy.

Recently, several THz pump-probe studies reveal semiconducting-to-metallic photoconductivity crossover in doped graphene[33-35]. These observations are accounted for the changes of Drude weight and transport scattering time by the doping level. The enhanced optical absorption of graphene in the tunable THz cavity could provide a new platform to elucidate non-ideal Drude response of graphene at high doping levels. Due to the frequency dependence of the optical conductivity, the maximum absorbance decreases with frequency. By combining this frequency dependence with the direct measurement of Fermi energy, we can extract the transport scattering time and its dependence on Fermi energy. In Figure 5a, we plot the normalized resonance absorbance, which is proportional to the real part of the normalized optical conductivity, $\sigma(\omega)/\sigma_{DC} = i/(\omega + i\tau^{-1})$, against the frequency for varying Fermi energy between 0.3 to 1.1 eV. Although at low doping concentration, THz response of graphene can be modeled with Drude model with constant scattering time, at high doping levels, however the scattering rate changes with Fermi energy. The transport scattering time has two contributions associated with the long-range charge impurity scattering ($\tau_c$) and short-range disorder scattering ($\tau_s$) as ($\tau^{-1} = \tau_c^{-1} + \tau_s^{-1}$).[26, 36] These scattering mechanisms scale differently with the Fermi energy. For short-range scattering, the scattering rate is proportional to $E_F$, however, for long-range scattering, the scattering rate is inversely proportional to $E_F$. Our results show that, as the Fermi energy increases, the absorbance decays slower with the increasing frequency indicating a smaller scattering time. Using the Drude model, we extracted the Fermi energy dependence of the total scattering time (Figure 5b). Around charge neutrality point, the scattering time is close to 100 fs and decreases down to 50 fs at Fermi energies of 1.1eV. These results revels the cylindrical with diameter of 2.7 cm. (c) THz reflectivity spectrum from the curved surface at different bias voltages.

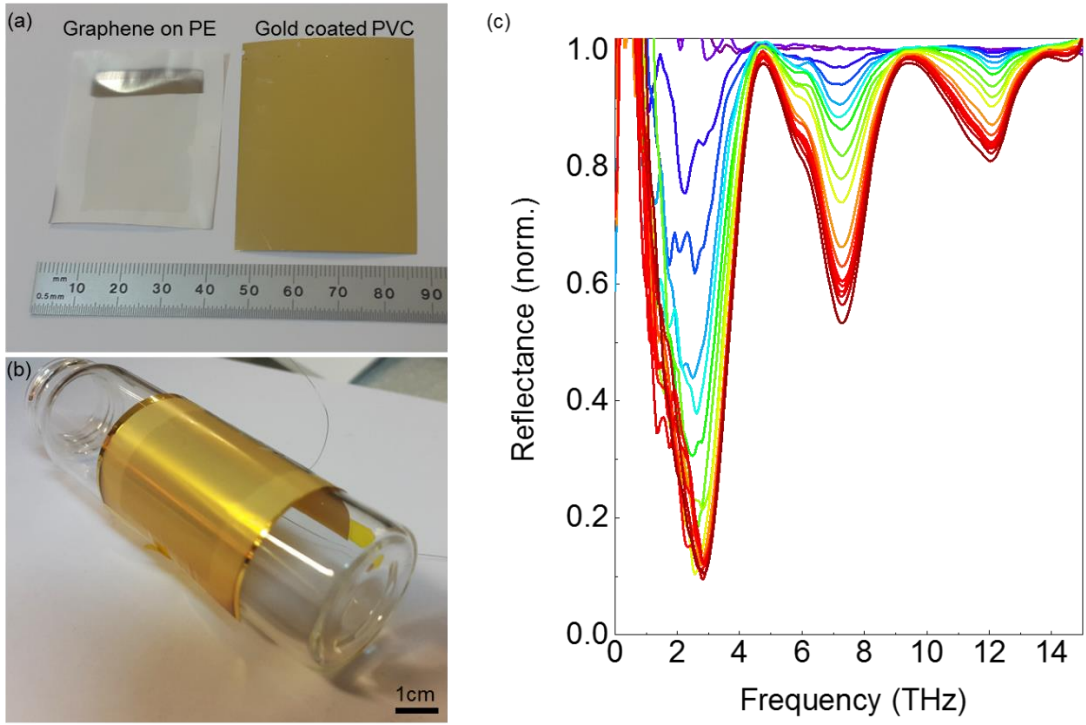

**Figure 6: Flexible active THz surfaces**: (a) Photograph of the large area graphene on PE membrane and gold coated PVC substrate. (b) The fabricated device rolled around a glass

To show the promises of our approach, we demonstrate flexible active THz surfaces as the application part of our work. The tunable coherent absorption of THz radiation can lead new types of active THz devices. Conventional THz devices are rigid which prevents realization of flexible THz components. The atomic thickness of graphene together with the simple device geometry allows us to fabricate a tunable THz cavity on a flexible polymer substrate. Figure 6a shows the photograph of large area graphene (2.5 x 3.0 $cm^2$) on porous PE membrane and gold coated PVC substrate. After injecting ionic liquid into the PE membrane (20 µm thick), we placed it on the gold coated PVC substrate and rolled the device around a glass cylindrical with diameter of 2.7 cm (Figure 6b). We measured the variation of the THz reflectivity from the curved surface. During the measurement the beam size is set to 6 mm in diameter. Similar to the rigid devices, we observed three resonances at 3, 7.2 and 12.2 THz (Figure 6c). The first resonance yields gate tunable absorption up to 95% at 2V bias voltage.

In conclusion, we report experimental observation of gate-tunable coherent perfect absorption of terahertz radiation in highly doped graphene. Our work has four novel parts. First, we developed an electrically tunable THz cavity using a THz transparent porous membrane soaked with ionic liquid electrolyte sandwiched between graphene and gold electrodes. In this device geometry gold electrode operates both as reflecting mirror and the gate electrode. Second, we observed the coherent perfect absorption of THz radiation in graphene. Ability to gate graphene up to 1eV Fermi levels in the THz cavity, allows us to observe critical coupling condition which yields absorption of 99%. Third, this novel device configuration allows direct measurement of Fermi energy and to elucidate the doping dependence of the transport scattering time which varies from 100 fs down to 50 fs as the Fermi energy changes between 0.2 to 1.1 eV. Finally, using these structures we demonstrated flexible active THz surfaces with voltage controlled THz reflectance. We anticipate that our work provide a developed device structure provides a new platform to study gate-tunable CPA would lead to efficient active THz components such as tunable THz mirrors and modulators.

**Acknowledgement**: This work was partially supported by the Scientific and Technological Research Council of Turkey (TUBITAK) grant no 114F379 and the European Research Council (ERC) Consolidator Grant ERC – 682723 SmartGraphene. N.K. acknowledges TUBITAK-BIDEB 2215 scholarship program.

**Supporting Information:**

Transmission line models for the free standing graphene and the active device, THz transmittance of the PE membrane, voltage controlled IR reflectivity from the device, variation of the center frequency and width of the resonance with the bias voltage, voltage controlled reflectivity of the device characterized by 0.368 THz continuous THz source.